\documentclass[journal,twocolumn]{IEEEtran}
\usepackage{epsfig,makeidx,color,subfigure}
\usepackage{amsmath,amssymb,bbm}
\usepackage{cite,graphicx,lipsum}
\usepackage{enumerate}
\usepackage{empheq}
\usepackage[switch,pagewise]{lineno}
\usepackage{hyperref}
\hypersetup{
        colorlinks = true,
        citecolor=blue,
}
\def\cM{{\cal M}}

\def\rT{{\rm T}}

\def\uP{{\mathbb P}}

\def\uE{{\mathbb E}}

\newtheorem{mylemma}{\bf Lemma} 

\def\be{ \begin{equation} }
\def\ee{ \end{equation} }
\def\bea{ \begin{eqnarray} }
\def\eea{ \end{eqnarray} }

\def\ba{{\bf a}}

\def\bm{{\bf m}}

\def\bR{{\bf R}}

\def\b0{{\bf 0}}

\def\cA{{\cal A}}
\def\cC{{\cal C}}

\def\cB{{\cal B}}

\def\cN{{\cal N}}

\def\sSINR{{\rm SINR}}

\ifCLASSOPTIONonecolumn
  \newcommand{\figwidth}{0.40\columnwidth}
\else
  \newcommand{\figwidth}{0.80\columnwidth}
\fi
\usepackage{graphicx}
\usepackage{booktabs}
\usepackage{graphicx}
\usepackage[normalem]{ulem}

\begin{document}

\title{On Throughput Bounds of NOMA-ALOHA}

\author{Jinho Choi\\
\thanks{The author is with
the School of Information Technology,
Deakin University, Geelong, VIC 3220, Australia
(e-mail: jinho.choi@deakin.edu.au).
This research was supported
by the Australian Government through the Australian Research
Council's Discovery Projects funding scheme (DP200100391).}}
\maketitle

\begin{abstract}
In this paper, we focus on the throughput of random access with power-domain non-orthogonal multiple access (NOMA) and derive bounds on the throughput. In particular, we demonstrate that the expression for the throughput derived in \cite{Yu21} is an upper-bound and derive a new lower-bound as a closed-form expression. This expression allows to find the traffic intensity that maximizes the lower-bound, which is shown to be the square root of the number of power levels in NOMA. Furthermore, with this expression, for a large number of power levels, we obtain the asymptotic maximum throughput that is increased at a rate of the square root of the number of power levels.
\end{abstract}


\ifCLASSOPTIONonecolumn
\baselineskip 23pt
\fi

\section{Introduction}

For the Internet of Things (IoT), we expect more and more sensors and devices to be connected to the Internet.
In order to support the connectivity of
a large number of devices and sensors (or things)
various connectivity technologies are proposed \cite{Ding_20Access}. Among those, 
machine-type communication (MTC) within cellular systems \cite{3GPP_MTC_18}
has been actively studied in the 5th generation (5G) to support a large number of devices over a large geographic area  for various applications 
(e.g., smart factory,
smart cities, and intelligent transportation) \cite{Dawy17} \cite{Bockelmann18}.

MTC differs from human-type communication (HTC)
in a number of ways. For example, MTC has a large number of devices, but their activity is sparse.
As a result, 
random access is preferable for uplink transmissions from devices due to low signaling overhead. As in \cite{Arouk14}  \cite{Choi16CL}, ALOHA is considered for MTC. 

Non-orthogonal multiple access (NOMA) has been extensively studied to improve the spectral efficiency of cellular systems, e.g., 5G, by exploiting the power domain with successive interference cancellation (SIC) at receivers \cite{Choi08} \cite{Saito13} \cite{Ding_CM} \cite{Choi17_ISWCS}. The application of NOMA to random access, especially ALOHA \cite{Abramson70}, is first considered in 
\cite{Choi_JSAC}, which is referred to as NOMA-ALOHA in this paper, and also in 
\cite{Balevi18} \cite{Choi19a} \cite{Tegos20}. 
In order to see the performance gain by NOMA, it would be necessary to 
find the throughput as a closed-form expression.
In \cite{Choi_JSAC}, a lower-bound on the throughput is found
(note that the equation after Eq. (14) in \cite{Choi_JSAC} is not correct as the second term on the right hand side (RHS) has to be $\frac{e^{-\lambda} \lambda^2}{2!}$). 
On ther other hand, it is claimed in 
\cite{Yu21} that an exact expression for the throughput is obtained.
 
In this paper, under the same conditions as in \cite{Yu21}, we show that the throughput in \cite{Yu21} is actually an upper-bound. As a result, any analysis and optimization based on the derived throughput may lead to optimistic views on the performance. In addition, if any attempts are made to stabilize NOMA-ALOHA systems based on an upper-bound on the throughput, they may unstabilize systems (because the traffic intensity can be increased to meet the throughput based on an upper-bound, which may lead to a too high traffic intensity that could overload the system).  
In this sense, a lower-bound could be useful. Note that although a lower-bound is found in \cite{Choi_JSAC}, it is not a compact expression due to a summation. 

We derive a low-bound on the throughput as a closed-form expression (under the same conditions as in \cite{Yu21}). This expression allows to find the optimal traffic intensity that maximizes the lower-bound, which is $\sqrt{Q}$, where $Q$ represents the number of power levels. In addition, for a large $Q$, it is shown that the maximum throughput of 
NOMA-ALOHA can be increased at a rate of $\sqrt{Q}$.

\subsubsection*{Notation}
Matrices and vectors are denoted by upper- and lower-case
boldface letters, respectively.
The superscript $\rT$
denotes the transpose.
For a set $\cA$, $|\cA|$ denotes the cardinality of $\cA$.
$\uE[\cdot]$ and ${\rm Var}(\cdot)$
denote the statistical expectation and variance, respectively.
$\cC \cN(\ba, \bR)$
represents the distribution of
circularly symmetric complex Gaussian (CSCG)
random vectors with mean vector $\ba$ and
covariance matrix $\bR$.

\section{System Model}

Suppose that there are a number of users for uplink transmissions to a base station (BS).
Due to users' sparse activity, a small number of users are active at a time.
Denote by $K$ the number of active users in a slot. Then, the received signal at the BS is given by \cite{TseBook05} 
\be 
r = \sum_{k = 1}^K h_k \sqrt{P_k} s_k + n,
\ee 
where $h_k$ is the channel coefficient from active user $k$
to the BS, $P_k$ and $s_k$ are the transmit power and signal of active user $k$, respectively, and $n \sim \cC \cN (0, \sigma^2)$ is the background noise. Here, $\sigma^2 = {\rm Var}(n)$.

Throughout this paper, it is assumed
that the channel state information (CSI) is known to users.
In time division duplexing
(TDD) mode, the BS can send a beacon signal at the beginning
of a time slot to synchronize uplink transmissions.
This beacon signal can also be used
as a pilot signal to allow each user to estimate the CSI.
As in \cite{Choi_JSAC}, there are pre-determined $Q$ power levels
that are denoted by 
\be
v_1 > \ \ldots \ > v_Q > 0.
	\label{EQ:vv}
\ee
We now assume that
an active user, say user $k$, can randomly choose
one of the power levels, say $v_q$, for 
random access. Then, the transmission power
can be  decided as
\be 
P_k = \frac{v_q}{\alpha_k},
	\label{EQ:CI}
\ee
where $\alpha_k = |h_k|^2$ is the channel (power) gain from
user $k$ to the BS, so that the
received signal power becomes $v_q$.
Assuming that the variance
of the background noise is normalized, i.e., $\sigma^2 = 1$,
if there are no other active users, the signal-to-noise ratio
(SNR) at the BS becomes
$v_q$.

Suppose that each power level in
\eqref{EQ:vv} is decided as follows:
\be
v_q = \Gamma (V_q + 1),
	\label{EQ:vV}
\ee
where $\Gamma$ is the target signal-to-interference-plus-noise ratio (SINR) for successful decoding and $V_q = \sum_{m=q+1}^Q v_m$
with $V_Q = 0$. As shown in \cite[Eq. (8)]{Choi_JSAC}, it can be shown that 
\be
v_q = \Gamma (\Gamma +1 )^{Q-q}.
	\label{EQ:vl}
\ee
These
multiple power levels, $\{v_1, \ldots, v_Q\}$, can be viewed as multiple channels that can be randomly selected by active users  when NOMA is applied to random access.  

For example,
let $Q = 4$ and $\Gamma = 2$. From \eqref{EQ:vl}, we have
\begin{align}
    v_4 & = \Gamma = 2, \  v_3  = \Gamma (1+\Gamma) = 6 \cr
    v_2 & = \Gamma (1+\Gamma)^2 = 18, \ \mbox{and} \ 
    v_1 = \Gamma (1+\Gamma)^3 = 54.
    \label{EQ:ex_vs}
\end{align}
If there exists at most one active user at each power level, all the signals from active users can be decoded. To see this, suppose that there is one active user at each power level. Then, the SINR for the active user who chooses $v_1$ becomes $\frac{v_1}{V_1 + 1} = \frac{v_1}{v_2+v_3+v_4+1} = \frac{54}{27} = \Gamma = 2$,  and the SINRs of the other active users are also $\Gamma = 2$. For another example, suppose that there is no active user at the 3rd power level. Then, the SINR for the active user who chooses $v_1$ is $\frac{v_1}{v_2+v_4+1} = \frac{54}{21}$, which is 
greater than $\Gamma$, and the SINRs of the other active users are greater than or equal to $\Gamma$.

\section{Throughput of NOMA-ALOHA with Multiple Channels}

In this section,
as in \cite{Choi_JSAC}, NOMA is applied to multichannel ALOHA to increase the throughput. We find bounds on the throughput and compare with known results in \cite{Choi_JSAC} \cite{Balevi18} \cite{Yu21}.

Throughout the paper, the throughput is defined as the average number of packets per slot that can be successfully decoded at the BS.
In order to find the throughput, we consider the following assumptions.
\begin{itemize}
    \item[{\bf A1})] Each active user can choose one of $L$ channels and one of $Q$ power levels uniformly at random.
    \item[{\bf A2})] The number of active users choosing channel $l$, denoted by $K_l$ follows the Poisson distribution with mean $\lambda$. As a result, the total number of active users, $K = \sum_{l=1}^L K_l$, follows the Poisson distribution with mean $\Lambda = L \lambda$, and the number of active users choose channel $l$ and power level $q$ follows the Poisson distribution with mean $\frac{\lambda}{Q}$.
\end{itemize}


\subsection{An Upper-Bound}

As discussed in \cite{Choi_JSAC} and \cite{Yu21}, the BS is to subsequently decode the signals at power levels 1 to $Q$ using SIC.
Under the assumptions of {\bf A1} and {\bf A2}, it has been claimed in \cite{Yu21} that an exact throughput is obtained as follows:
\be 
\eta = L \frac{\lambda}{Q} e^{-\frac{\lambda}{Q}}
\sum_{q=0}^{Q-1} \left(
\left(1+\frac{\lambda}{Q} \right)
e^{-\frac{\lambda}{Q}}
\right)^{q}.
    \label{EQ:Yu}
\ee 
To find the throughput in \eqref{EQ:Yu}, 
due to the symmetric conditions according to the assumptions of {\bf A1} and {\bf A2}, we can consider one channel, say channel $l$. 
Let $M_{l,q}$ represent the number of the active users choose the $l$th channel with power level $q$ (note that $K_l = \sum_{q=1}^Q M_{l,q}$). For successful decoding at power level $1$, we need $M_{l,1} = 1$ so that no power collision happens. In addition, 
for successful decoding at power level $2$, we need $M_{l,2} = 1$ as well as $M_{l,1} \in \{0,1\}$ (i.e., no signal or one signal so that SIC can be performed after successful decoding). Let $a$ and $b$ be the probabilities that $M_{l,q} = 0$ and $1$, respectively. Due to the the symmetric conditions, $a$ and $b$ are independent of $l$ and $q$.
Then, the average number of signals that can be successfully decoded in channel $l$ is given by
\be
\eta_l = \sum_{q=1}^Q b (a+b)^{q-1}
= b \sum_{q=0}^{Q-1} (a+b)^q.
    \label{EQ:el}
\ee 
Here, according to Assumption of {\bf A2}, we have
\begin{align}
a & = \Pr(M_{l,q} = 0) = e^{-\frac{\lambda}{Q}} \cr 
b & = \Pr(M_{l,q} = 1) = \frac{\lambda}{Q}e^{-\frac{\lambda}{Q}}. 
    \label{EQ:ab}
\end{align} 
By substituting \eqref{EQ:ab} into \eqref{EQ:el}, we have \eqref{EQ:Yu}, 
which can also be expressed as
\be 
\eta = L \eta_l = L \frac{\lambda}{Q} e^{-\frac{\lambda}{Q}}
\frac{1 - (a+b)^Q}{ 1- (a+b)}.
    \label{EQ:eLe}
\ee 
Unfortunately, this approach does not provide an exact throughput, but an upper-bound.
To see this, consider the case that $\bm = [M_{l,1}
\ \ldots \ M_{l,Q}]^\rT = [1 \ 1 \ 3 \ 1]^\rT$ with $Q = 4$ and $\Gamma = 2$. 
According to the way described above, the signals at power levels 1 and 2 should be decodable. However, from \eqref{EQ:ex_vs}, the SINR to decode the signal at power level 1 is given by
$$ 
\sSINR = \frac{v_1}{ v_2 + 3 v_3 + v_4 + 1} = \frac{54}{18+18+ 2+ 1} = 
1.3846,
$$ 
which is less than $\Gamma = 2$. Thus, there is no decodable signal at all. As a result, \eqref{EQ:Yu} includes the case that the BS is unable to decode and leads to an upper-bound on the throughput.

\subsection{A Lower-Bound}

Let
\be 
\gamma_{l,q}(\bm_{l,q+1}) = \frac{v_q}{M_{l,q+1} v_{q+1} + \ldots + M_{l,Q} v_Q + 1},
\ee 
where $\bm_{l,q+1} = [M_{l,q+1}, \ldots, M_{l,Q}]^\rT$, which is the SINR of the signal at power level $q$ when all the signals at power levels $1$ to $q-1$ are removed by SIC. Clearly, $\bm_l = \bm_{l,0}$ (and $\bm_{l,q+1}$ is a subvector of $\bm_l$). Define a set of subvectors of $\bm_l$ as
\be 
\cM_{-q} = \{ \bm_{l,q+1}: \ \gamma_{l,q}(\bm_{l,q+1}) \ge \Gamma \}
\ee 
and denote by $\bar q$ the last power level without power collisions yet, which is given by
$$
\bar q (\bm_l) = \min_{1 \le q \le Q} \{q: \ M_{l,q} \ge 2 \} - 1.
$$
If all $M_{l,q} \le 1$, $\bar q (\bm_l) = Q$.
Noting that SIC cannot be carried out after $\bar q$,
for a given $\bm_l$, the number of active devices that can successfully transmit in NOMA-ALOHA is 
\be 
\eta(\bm_l) = | \{M_{l,q} = 1: \bm_{l,q+1} \in \cM_{-q}, q = 1, \ldots, \bar q (\bm_l)\}|.
\ee

For example, as before, let $Q = 4$ and $\Gamma = 2$. Suppose that
$\bm_l = [1\ 0 \ 4 \ 1]^\rT$. Then, we have $\bar q = 2$.
Since $M_{l,1} = 1$ and
\begin{align*}
\bm_{l,2} & =[0 \ 4 \ 1]^\rT \in
\cM_{-1} \cr
& \ \ = \left\{ \bm_{l,2}: \ \frac{54}{18M_{l,2} +6 M_{l,3} +2 M_{l,4} + 1} \ge 2 \right\}
\end{align*}
from \eqref{EQ:ex_vs},
we have $\eta(\bm_l) = 1$.


\begin{mylemma} \label{L:1}
If $\bm_l \in \cB^Q$, where $\cB = \{0,1\}$, i.e., $\bm_l$ is a $Q\times 1$ vector consisting of 0 or 1,
then $\eta(\bm_l) = ||\bm_l||_1$.
\end{mylemma}
\begin{IEEEproof}
According to \eqref{EQ:vl}, for any $\bm_l \in \cB^Q$, the SINR is greater than or equal to $\Gamma$, i.e., $\gamma_{l,q}
(\bm_{l,q+1}) \ge \Gamma$. Thus, all the active devices with $\bm_l \in \cB^Q$ are decodable. This means that the number of the successfully decoded signals is equal to $||\bm_l||_1 = \sum_{q=1}^Q M_{l,q}$.
\end{IEEEproof}

Based on Lemma~\ref{L:1}, we can find a lower-bound on the throughput. 
As explained above, although $\bm \notin \cB^Q$, the BS can decode some signals (e.g., the BS can decode one signal if $\bm_l = [1 \ 0 \ 4 \ 1]^\rT$. However, considering only $\bm \in \cB^Q$, we can have a lower-bound as follows.

According to Lemma~\ref{L:1} and the assumptions of {\bf A1} and {\bf A2},
the conditional throughput for $\bm \in \cB^Q$ is given by
\begin{align}
\nu_l & =  \uE[ \eta(\bm_l) \,|\, \bm_l \in \cB^Q ] \cr
& = \sum_{q=1}^Q q \Pr(||\bm_l||_1 = q \,|\, \bm_l \in \cB^Q ) \cr 
& = \sum_{q=1}^Q q \binom{Q}{q} \bar b^q \bar a^{Q-q} = Q \bar b,
\end{align} 
where $\bar a$ and $\bar b$ are $\frac{a}{a+b}$ and $\frac{b}{a+b}$, respectively. Let $\uP_\cB = \Pr(\bm_l \in \cB^Q)$.
Then, the throughput (per channel) is lower-bounded as
\begin{align}
\eta_l  = \nu_l \uP_\cB + \uE[ \eta(\bm_l) \,|\, \bm_l \notin \cB^Q ](1 - \uP_\cB) 
\ge \nu_l \uP_\cB, 
\end{align}
i.e., the lower-bound is obtained by ignoring the second term
on the RHS.
Since the number of active devices at each power level is independent,
we have
\be 
\uP_\cB = \prod_{q=1}^Q \Pr (M_{l,q} \in \{0,1\}) = (a + b)^Q.
\ee
Finally, the following lower-bound can be obtained:
\begin{align} 
\eta_l & \ge Q b (a+b)^{Q-1} \cr 
& = \lambda  e^{- \frac{\lambda}{Q}}
\left(\left(1+\frac{\lambda}{Q} \right)
e^{-\frac{\lambda}{Q}}
\right)^{Q-1}.
    \label{EQ:LB}
\end{align} 

\begin{mylemma}
The maximum of the lower-bound in \eqref{EQ:LB} happens
at $\lambda = \sqrt{Q}$. The maximum of the lower-bound is
\be 
w(Q) = \sqrt{Q} e^{-\sqrt{Q}} 
\left( 1+\frac{1}{\sqrt{Q}}\right)^{Q-1}.
    \label{EQ:wQ}
\ee 
\end{mylemma} 
\begin{IEEEproof}
For convenience, let $x = \frac{\lambda}{Q}$. Then, the lower-bound in \eqref{EQ:LB} is given by
\be
f(x) = Q x e^{-x} \left( (1+x) e^{-x} \right)^{Q-1}.
\ee 
Then, the derivative is
\be 
\frac{d f(x)}{dx}
= Q \left( (1+x) e^{-x} \right)^{Q-2} e^{-2x} (1 - Q x^2),
\ee 
which is positive when $x < \frac{1}{\sqrt{Q}}$ and negative $x > \frac{1}{\sqrt{Q}}$. Thus,
$f(x)$ is a unimodal function that has the maximum when $x = \frac{1}{\sqrt{Q}}$. That is, the lower-bound in \eqref{EQ:LB} has a unique maximum at $\lambda = \sqrt{Q}$. With $\lambda = \sqrt{Q}$, the maximum of the lower-bound as a function of $Q$ in \eqref{EQ:wQ} can be obtained.
\end{IEEEproof}

For a large $Q$, from \eqref{EQ:wQ}, using $1+x \approx e^{x}$
for $|x| \ll 1$, we have
\be
w(Q) \approx \sqrt{Q} e^{-\sqrt{Q}} e^{\frac{Q-1}{\sqrt{Q}} }
= \sqrt{Q} e^{-\frac{1}{\sqrt{Q}}} \approx \sqrt{Q}-1.
\ee 
This shows that the maximum throughput of 
NOMA-ALOHA can be increased at a rate of $\sqrt{Q}$.

In summary, from \eqref{EQ:el} and \eqref{EQ:LB},
the throughput per channel, $\eta_l$,
is bounded as
\begin{align}
& \frac{\lambda}{Q} e^{-\frac{\lambda}{Q}}
\sum_{q=0}^{Q-1} \left(
\left(1+\frac{\lambda}{Q} \right)
e^{-\frac{\lambda}{Q}}
\right)^{q} \ge \eta_l \qquad  \qquad \cr
& \qquad \qquad \ge \lambda  e^{- \frac{\lambda}{Q}}
\left(\left(1+\frac{\lambda}{Q} \right)
e^{-\frac{\lambda}{Q}}
\right)^{Q-1}.
    \label{EQ:bb}
\end{align}
Note that the 
the bounds in \eqref{EQ:bb} become the exact one, i.e.,
$\lambda e^{-\lambda}$, when $Q = 1$ (which is the case of conventional ALOHA). Furthermore, the
lower-bound  is exact when $Q = 2$, which is
$\lambda e^{-\lambda} + \frac{\lambda^2}{2} e^{-\lambda}$, as discussed in \cite{Choi_JSAC}. However, with $Q = 2$, the upper-bound, which is $\frac{\lambda}{2} e^{-\frac{\lambda}{2}} \left(1 + \left(1+\frac{\lambda}{2} \right)
e^{-\frac{\lambda}{2}} \right)$, is not exact.

\section{Simulation Results}

In this section, we present simulation results of throughput under the assumptions of {\bf A1} and {\bf A2}.
Since the total throughput is linearly proportional to the throughput per channel (as shown in \eqref{EQ:eLe}), we only show the throughput per channel, $\eta_l$,
with the bounds in \eqref{EQ:el} and \eqref{EQ:LB}. 

\begin{figure}[thb]
\begin{center} 
\subfigure[$Q = 2$]{\label{fig 0 ay}
\includegraphics[width=0.4\textwidth]{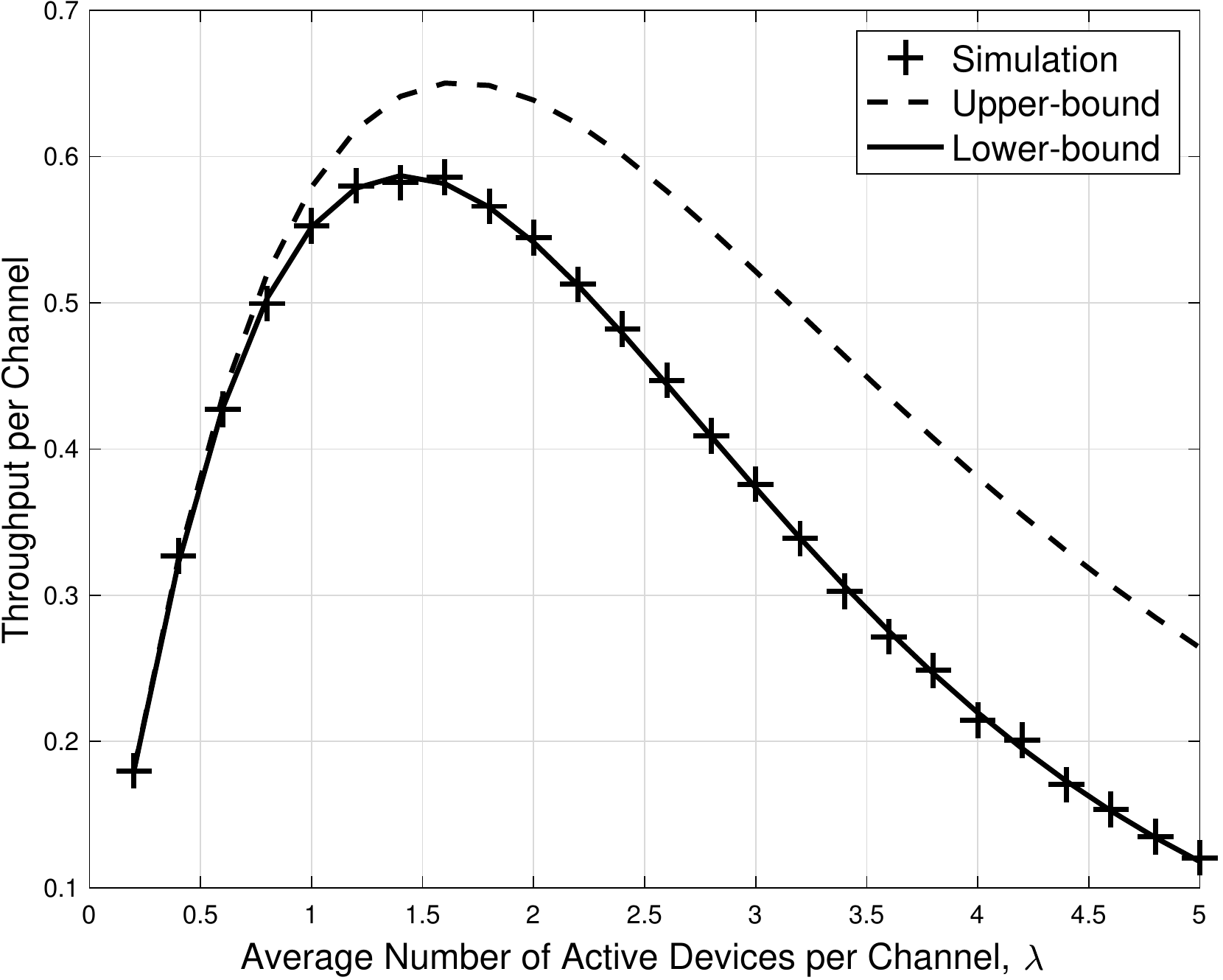}}
\subfigure[$Q = 4$]{\label{fig 0 by}
\includegraphics[width=0.4\textwidth]{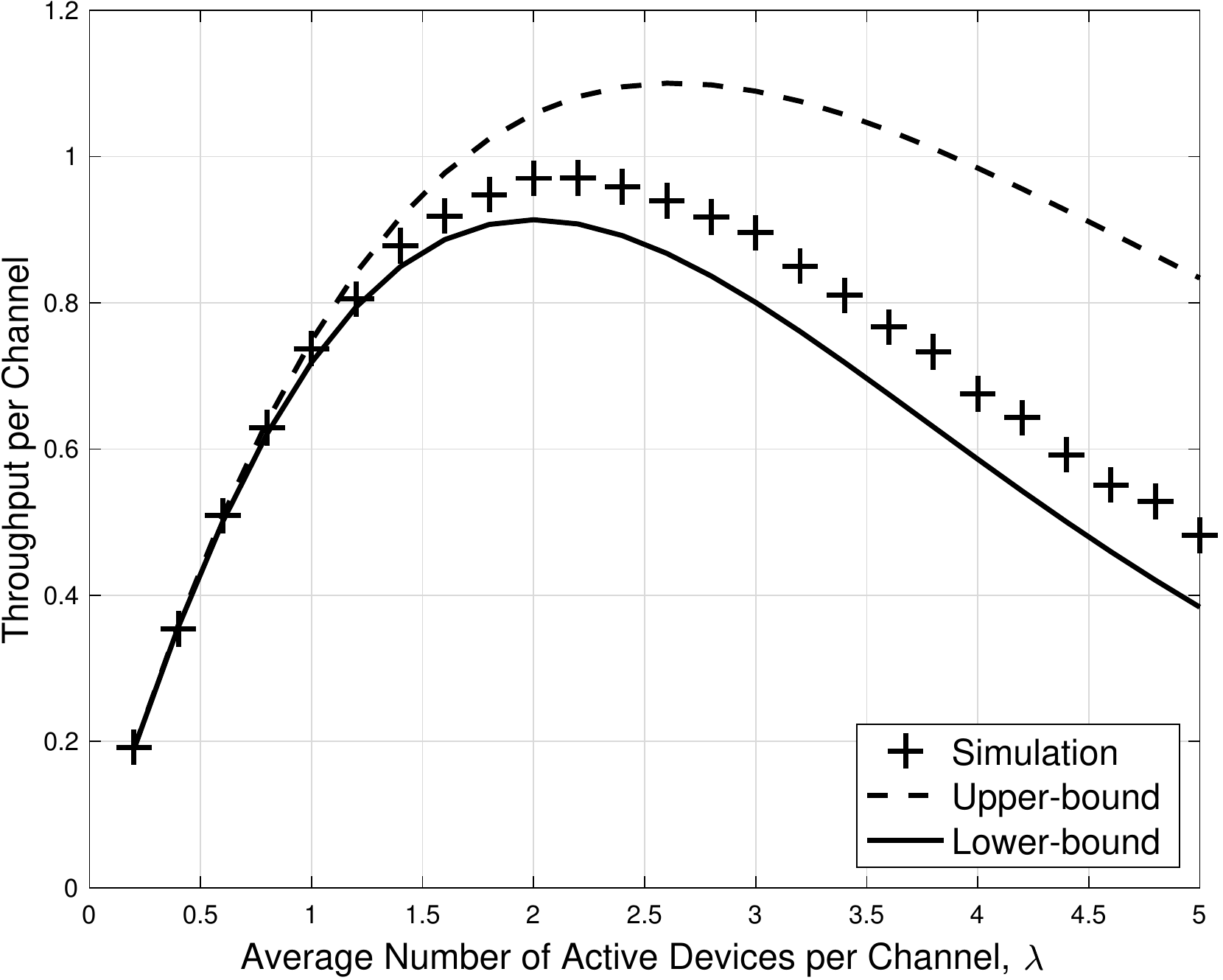}} 
\end{center}
\caption{Throughput as a function of traffic intensity, $\lambda$, when
$\Gamma = 4$ dB: (a) $Q = 2$; (b) $Q = 4$.}
        \label{Fig:plt1}
\end{figure}

Fig.~\ref{Fig:plt1}
shows the throughput per channel as a function of the traffic intensity or average number of active users per channel, $\lambda$, when $\Gamma = 4$ dB and $Q \in \{2, 4\}$.
While \eqref{EQ:LB} becomes the exact expression for the throughput when $Q = 2$ as shown in Fig.~\ref{Fig:plt1} (a), it becomes a lower-bound when $Q > 2$ as shown in Fig.~\ref{Fig:plt1} (b). 
We can also see that \eqref{EQ:el} (the expression in \cite{Yu21}) is an upper-bound.

\begin{figure}[thb]
\begin{center}
\includegraphics[width=\figwidth]{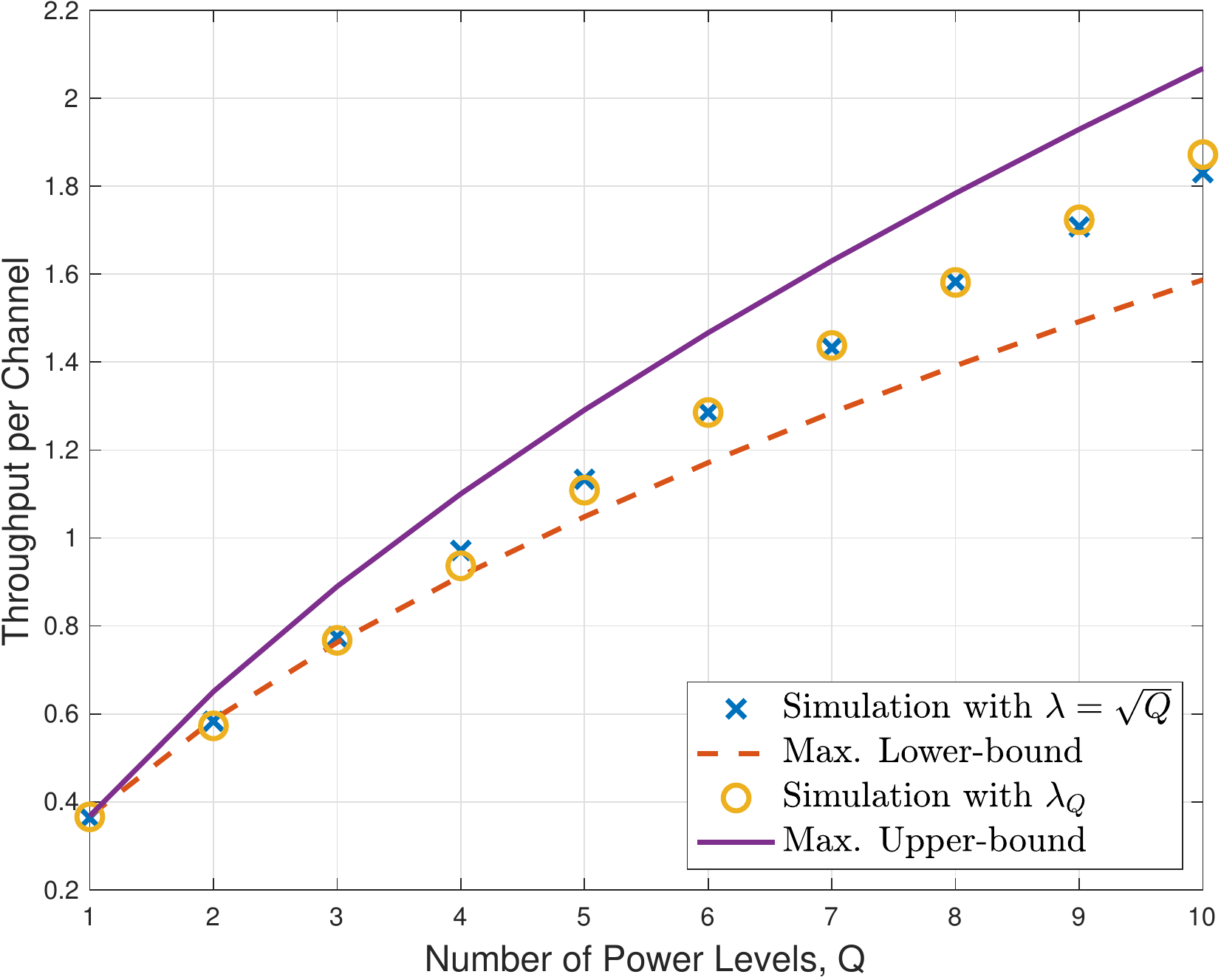}
\end{center}
\caption{Throughput as a function of the number of power levels, $Q$, when
$\Gamma = 4$ dB.}
        \label{Fig:plt3}
\end{figure}

Let $\lambda_Q$ be the value of $\lambda$ that maximizes the upper-bound in \eqref{EQ:Yu}. As discussed earlier, $\lambda = \sqrt{Q}$ maximizes the lower-bound. With the two optimal values of $\lambda$, we run simulations for different values of $Q$ and present the results in Fig.~\ref{Fig:plt3} with $\Gamma = 4$ dB.
We can see that the lower-bound agrees with the simulation results when $Q$ is small (i.e., $Q \le 4$). However, as $Q$ increases, the lower-bound becomes loose as the upper-bound. 

As discussed in \cite{Choi_JSAC}, a large $Q$ is impractical since a high transmit power is required. Thus, the lower-bound that is tight for a small $Q$ would be preferable to the upper-bound in \eqref{EQ:el}.

\section{Concluding Remarks}

We showed that the throughput expression derived in \cite{Yu21} is an upper-bound and found a lower-bound on the throughput as a closed-form expression. Based on this expression, it was shown that the throughput can be maximized if $\lambda = \sqrt{Q}$ and be increased at a rate of $\sqrt{Q}$. Since the transmit power increases exponentially with $Q$, this result indicates that a small $Q$ is preferable over a large $Q$ in practice (say $Q =2$). 

While we mainly focused on the throughput in this paper, various issues can be addressed using the derived expression. For example, user barring algorithms can be derived using the lower-bound. In addition, a stability analysis can be carried out based on the lower-bound. Those will be studied as further research topics in the future.

\bibliographystyle{ieeetr}
\bibliography{mtc}

\end{document}